\documentclass[aip,apl,numerical,amsmath,amssymb,graphicx,reprint,citeautoscript]{revtex4-1}

\usepackage{graphicx}
\usepackage{dcolumn}
\usepackage{bm}
\usepackage{epstopdf}

\begin{document}


\title{Lateral spatial switching of excitons using vertical electric fields in semiconductor quantum rings } 




\author{P G McDonald}
\affiliation{Physics Department, School of Engineering and Physical Sciences, SUPA, Heriot-Watt University, Edinburgh EH14 4AS, United Kingdom}

\author{J Shumway}
\affiliation{Department of Physics and Astronomy, Arizona State University, Tempe, Arizona 85287-1504, USA}

\author{I Galbraith}
\email[]{I.Galbraith@hw.ac.uk}
\affiliation{Physics Department, School of Engineering and Physical Sciences, SUPA, Heriot-Watt University, Edinburgh EH14 4AS, United Kingdom}



\begin{abstract}
We study the response of exciton complexes in semiconductor quantum rings to
vertical electric fields, using path integral quantum Monte Carlo simulations. 
The interaction of a vertical applied field and the piezoelectric fields of the ring with
strongly-correlated excitonic complexes
switches excitons between two different lateral locations within the ring . 
This control should be observable through polarizability and dipole measurements,
and, for biexcitons, an energy shift beyond the normal Stark shift.
\end{abstract}


\maketitle 

Semiconductor nanostructures 
allow for the tailoring of their optical and electronic properties by manipulation of their
size, shape, and composition profile.
A widely-studied class of nanostructures are InGaAs/GaAs self-assembled quantum dots,
which have found a wide range of applications from single photon sources 
to quantum information processing.\cite{Imamog_prl_1999,Yuan_science_2002}
Beyond such passive tailoring of properties during growth,
active control of nanostructure properties through external fields,
especially switching behavior, is highly desirable.

Under proper annealing conditions, quantum dots undergo 
a significant migration of In outwards from the center of the dot, forming a ring-shaped 
nanostructure with a crater at the center 
\cite{Lorke_JJAP_2001, offermans_APL_2005, granados_APL_2005,%
Baranwal_PRB_2009, mlakar_APL_2008, granados_APL_2003, garcia_APL_1997,%
pettersson_physicae_2000}. Previous theoretical and experimental work has show 
that quantum rings behave differently from their quantum dot cousins
\cite{Warburton_PRB_2002, alen_PRB_2007, Galbraith_pss_2002}, with 
distinct  but equally rich electronic and optical properties. 
These unique properties stem from 
the GaAs barrier material at the ring's core, which significantly modifies the confinement potential 
and---at the heart of the phenomena we report here---creates 
a strain profile that is unique to rings.
An important difference that has been previously identified is the distribution
of the piezoelectric field, which is seen to play a much more important 
role in rings than in dots \cite{Baker_PRB_2004}. Unlike in dots, where the
majority of the piezoelectric potential sits outside the electrically active part of the
structure---leaving the electron and hole ground states mostly unaffected---the 
strained central core of GaAs material in a 
quantum ring induces large piezoelectric potentials within the confining structure itself.
These piezoelectric fields break the rotational symmetry of the ring,
vertically separate the electron and hole, and induce 
localization.

\begin{figure}[b]
\includegraphics[width=8cm]{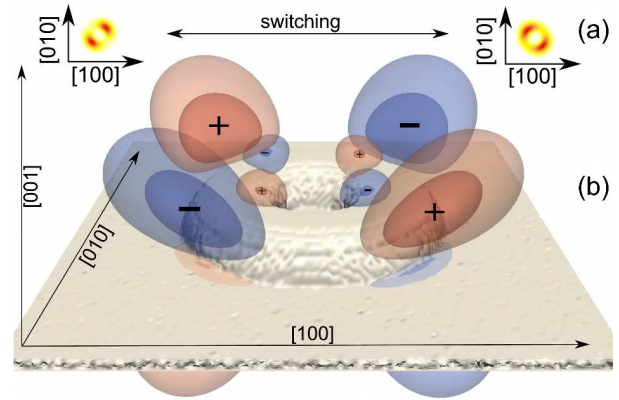}
\caption{ (Color online) (a) Hole density for two different  localizations of excitons.
(b) Illustration of negative (blue) and positive (red) piezoelectric potential superimposed 
upon the structure of our model ring.  
 Each lobe is represented by two isosurfaces of 
$\pm$38 meV and $\pm$54 meV.}
\label{fig1}
\end{figure}

We theoretically investigate quantum rings and predict that the interplay of piezoelectric 
and strain effects  with electron-hole Coulomb interactions, will enable a \emph{lateral} switching 
of the probability distribution of the exciton when a \emph{vertical} electric field is applied. 
As illustrated in Fig.~\ref{fig1}(a), the exciton distribution in the ring can rotate 90$^\circ$.
Such two-state behavior resembles  a pair of
electrons in quantum dot cellular automata (QCA) cell~\cite{Snider_JJAP_1999},
and we address connections to QCA later in the discussion.
Further we show that this switching should be experimentally observable by a change 
in the lateral polarizability.

{\em Method---}%
Based on the XTEM image of a quantum ring observed by Lin et al \cite{Lin_APL_2009},
we construct a model ring that is 6~nm at its maximum height with an inner radius 5 nm 
and  outer radius 20 nm, made from a 50$\%$
(In,Ga)As random alloy, sitting upon a thin 30$\%$ In wetting layer. 
We carried all simulations at a temperature of 10 K.

We use a two band effective mass model Hamiltonian with conduction and valence bands 
derived from the strain profile of an atomistic model of the nanostructure
using the valence force field method~\cite{harowitz_JLTP_2005}. 
To simplify our analysis and presentation, we created eight atomistic models of the same 
ring, with different realizations of the random alloy, and present the average of
these eight rings, with ideal C$_{2v}$ symmetry imposed.
The piezoelectric potential is computed from the strain field and are included to second order 
following the work of Bester et al \cite{Bester_PRL_2006,*Bester_PRB_2006} and others \cite{schliwa_PRB_2007,Min_CPB_2009}.
Path integral quantum Monte Carlo \cite{Ceperley_RMP_1995, harowitz_JLTP_2005,pi-qmc} is then used to solve the system for exctions and biexcitons. 
This method properly treats quantum correlation---which is crucial for proper description of excitons in rings---at realistic finite temperatures with arbitrary three-dimensional confining potentials. We obtain polarizabilites from path integral quantum Monte Carlo using linear response theory.


{\em Piezoelectric fields and switching behavior---}%
As can be seen in Fig. \ref{fig1},
the ring is sectioned into quarters by the piezoelectric field, 
which has C$_{2v}$ symmetry.
Fig.~\ref{fig1}(b) clearly shows the top eight lobes
in the piezoelectric potential, with an equal number of lobes (with opposite sign) directly 
beneath the ring, two of which can be seen at the front edge.
The larger lobes, on the outer edge of the ring,  resemble the piezoelectric potential
found in a quantum dot; the majority of the field sits outside of the structure.
The extra, smaller lobes within the GaAs core,
which sit directly above and below the quantum ring and 
penetrate the structure, are not found in dots. These extra lobes have a significant effect on 
the quantum ring's electronic structure.

\begin{figure}[b]
\includegraphics[width=8cm]{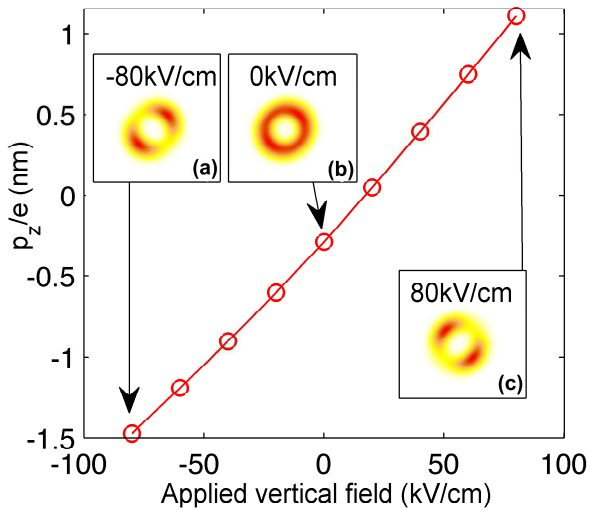}
\caption{Vertical dipole, $p_z$, for different vertical applied electric fields, shows changing of sign, and permanent dipole at zero field. Insets (a), (b) and (c) show mean electron probability distribution for -80kV/cm, 0 and 80kV/cm and demonstrates switching of the exciton around ring by changing sign of
vertical field}\label{fig2}
\end{figure}

An electric field in the vertical growth direction ([001]), perpendicular to the plane of the ring, 
polarizes the exciton in the growth direction, Fig.~\ref{fig2}.
The vertically polarized excitons are attracted
to the smaller piezoelectric potential lobes {\em in the GaAs core} (Fig.~\ref{fig1})
that align with the induced excitonic dipole.
Changing the direction of the vertical electric field will cause either the 
[110] or [1$\bar{1}$0]  direction to be lower in energy,
while the other diagonal higher in energy. 
This switching of the lateral localization with a changing of the vertical 
electric field direction can be seen in Fig. \ref{fig2}(a) and (c), which shows how the mean electron
probability distribution aligns along either [110] or [1$\bar{1}$0]
depending on the sign of the field.

\begin{figure}[t]
\includegraphics[width=8cm]{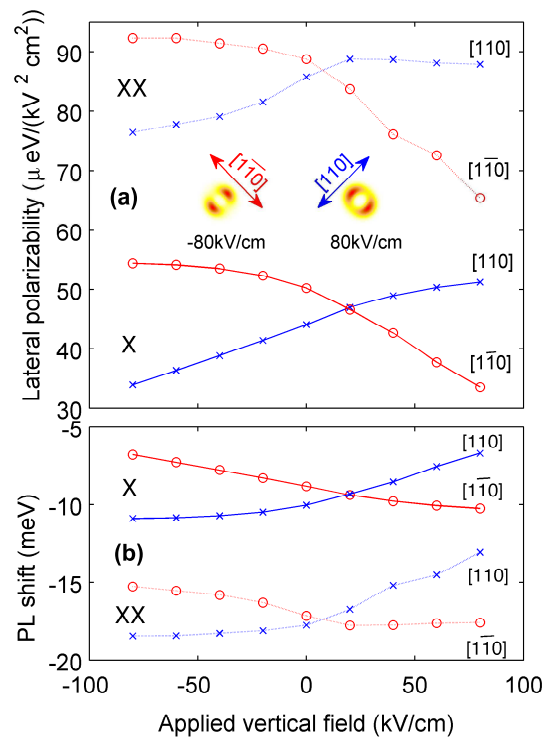}
\caption{(a)Lateral Polarizability of exciton and biexciton in the [110] and [1$\bar{1}$0]  direction against applied vertical electric field. 
Inset shows mean hole probability distribution for -80 kV/cm and +80kV/cm, with arrow indicating direction of strongest polarizability.
(b) PL shift vs.~vertical field with a lateral field of 20~kV/cm, with average parabolic shift removed for clarity. }
\label{fig3}
\end{figure}

{\em Observable effect on the lateral polarizability---}%
This switching should be experimentally visible by examining the in plane polarizability of the exciton and biexciton in the ring. As the vertical electric field is applied and the exciton localizes into the appropriate nodes as in  Fig. \ref{fig2} (a) and Fig. \ref{fig2} (c), there is a significant change in the polarizability of the complex. For example, a positive field causes
the exciton to localize in the [1$\bar{1}$0]  direction, resulting in an
 increase in the lateral polarizability tangential to the direction of confinement---the [110]
direction---as seen in Fig. \ref{fig3}. A negative field causes a 
larger polarizability in the [1$\bar{1}$0] direction.

From our electron-hole correlation data (not shown) we can see that throughout the switching 
of the applied vertical electric field the exciton stays bound, an increase in the field 
merely increases the separation between the electron and hole, increasing the vertical dipole as in Fig. \ref{fig2} as such the dipole can also be used as a experimental observable 
in conjunction with the polarizability e.g. it's effect on altering the recombination lifetime \cite{Wimmer_prb_2006}.
Due to the piezoelectric potential inverting its sign every
90$^\circ$, the applied fields will not only align the exciton along different 
diagonals with a switch in field direction but also be accompanied by a change 
in the direction of the dipole moment in the z-direction.

\begin{figure}[t]
\includegraphics[width=8cm]{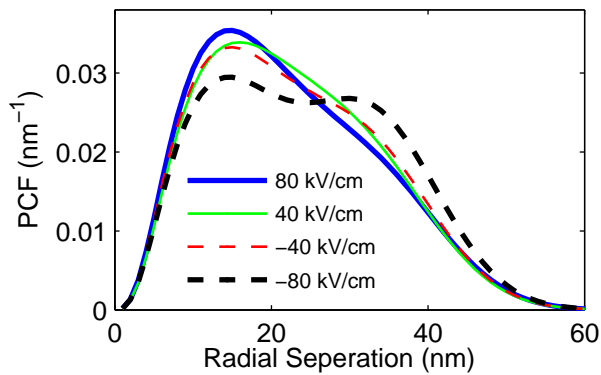}
\caption{ Pair Correlation Function (PCF) in a biexciton under various vertical electric fields, showing the partial dissociation of the biexciton
into two excitons. }
\label{fig4}
\end{figure}

The larger polarizability in [1$\bar{1}$0] than in [110] (as in Fig. \ref{fig3}) and the negative permanent dipole that exists in the quantum ring at zero field (as in Fig. \ref{fig2})  show that the exciton tends to align preferentially in the [110] direction at zero field, with the holes towards the bottom of the ring. As has been previously shown \cite{Warburton_PRB_2002}, the holes tend to stay away from areas of high biaxial strain, here located near the top of the quantum ring, whilst the electrons do not suffer such an issue and are more free to spread out. This explains the permanent dipole in the quantum ring. The preferential alignment and thus polarizability difference at zero field can again be put down to 
the piezoelectric effect, with the exciton aligning with the nodes which match the holes affinity towards the bottom of the ring, and so further lower the complex's energy.

{\em Structure of the biexciton---}%
For a biexciton complex, there is a subtle change from the case of the always bound exciton. The larger confinement for holes towards the bottom
of the ring will force the two holes together for a negative vertical electric field.
As the holes are forced closer together with a stronger field, this higher energy cost 
can counter the biexciton binding energy and cause it
to begin splitting into two excitons on either side of the ring. A positive electric field 
where the holes are towards the top of the ring, will tend to split the
biexciton more slowly as the extra confinement caused by the ring structure itself is 
now missing.
This is illustrated in Fig. \ref{fig4}, which shows the hole-hole separation 
at $\pm$40 kV/cm are similar,
however when increased to $\pm$80 kV/cm the negative field begins to split the 
biexciton, which can be seen as the emergence of a second peak at approximately 20nm, 
while the positive field continues to reduce the separation and no second peak is visible.
We suspect the difference in behavior between the positive and negative fields on the quantum 
ring in the case of the biexciton accounts for the more unusual polarizability seen 
for the biexciton seen in Fig. \ref{fig3}.

{\em Relation to QCA---}%
While the geometry of the exciton and biexciton localization and switching behavior
are reminiscent of electrons in four-dot QCA cells, the excitons are neutral objects.
Interactions between quantum rings, which are necessary for QCA 
circuits,\cite{Snider_JJAP_1999}
are dipole-dipole, hence very weak and only effective at low temperature,
perhaps in quantum information processing~\cite{Biolatti_PRL_2000}.

{\em Summary---}%
In conclusion, we have shown that it is possible to gain an extra degree of freedom
via the lateral control of an exciton complex in a quantum ring nanostructure by exploiting 
its unusual inbuilt piezoelectric field, and through the application of an external vertically applied electric field.
We also suggest how this may be experimentally detected through polarizability and dipole measurements.

P.G.M and I.G. acknowledge funding from EPSRC, Carnegie Trust, J.S from the NSF DMR 02-39819.

\bibliography{apl_ref}

\end{document}